\documentclass[aps,prd,twocolumn,showpacs,showkeys,superscriptaddress]{revtex4}

\usepackage{latexsym,amsmath}
\usepackage[dvips]{graphicx}

\begin{document}

%\preprint{arXiv:yymm.nnnn [gr-qc]}

\title{Note on cosmology of dimensionally reduced gravitational Chern-Simons}

\author{Wontae Kim}
\email[]{wtkim@sogang.ac.kr}
\affiliation{Department of Physics, Sogang University, Seoul 121-742, Korea}
\affiliation{Center for Quantum Spacetime, Sogang University, Seoul 121-742, Korea}

\author{Edwin J. Son}
\email[]{eddy@sogang.ac.kr}
\affiliation{Department of Physics, Sogang University, Seoul 121-742, Korea}

\date{\today}

\begin{abstract}
  We present cosmological solutions from the dimensionally reduced
  Chern-Simons term and obtain the smooth transition solution from the
  decelerated phase (AdS) to the accelerated phase (dS).
\end{abstract}

\pacs{04.60.Kz, 04.20.Jb}

\keywords{gravitational Chern-Simons, lower dimensional cosmology}

\maketitle

%%%%%%%%%%%%%%%%%%%%
%%  Introduction  %%
%%%%%%%%%%%%%%%%%%%%

A three-dimensional Einstein-Hilbert action has been dimensionally
reduced to two-dimensional lineal gravity theory~\cite{jt}, and
three-dimensional rotating black hole can be studied in the reduced
theory~\cite{ao,bdr} (for a review, see~\cite{gkv}).  On the other
hand, there has been much attention to the gravitational Chern-Simons
(GCS) term in the topologically massive gravity~\cite{djt} as a higher
derivative correction.  Recently, it has been shown that the
dimensional reduction from three-dimensional GCS action~\cite{gijp}
produces an interesting two-dimensional gravity model coupled to a
Maxwell field,
\begin{equation}
\label{action}
S_{CS} = -\frac{1}{8\pi^2} \int_{\mathcal{M}} d^2x \sqrt{-g} \left[ R f + f^3 \right],
\end{equation}
and this reduced GCS action allows kink solution~\cite{gijp,gk}, where
$f$ is the invariant field strength defined by
$f=-(1/2)\epsilon^{\mu\nu}F_{\mu\nu}$ and $\epsilon^{\mu\nu}$ is a
two-dimensional anti-symmetric tensor defined by
$\epsilon^{01}=1/\sqrt{-g}$.
In this Brief Report, we would like to
obtain cosmological solutions in this model.
%%%%%%%%%%%%%%%%%
%%  Solutions  %%
%%%%%%%%%%%%%%%%%
Varying 2d reduced GCS action~(\ref{action}), we obtain the equations
of motion as
\begin{align}
& \left[ \nabla_\mu \nabla_\nu f - g_{\mu\nu} \left( \Box f + f^3 + \frac12 R f \right) \right] = 0, \label{eq:met} \\
& \epsilon^{\mu\nu} \partial_\mu \left[ R + 3 f^2 \right] = 0. \label{eq:field} 
\end{align}
Then, solving Eq.~(\ref{eq:field}), we get
\begin{equation}
R + 3 f^2 = \alpha, \label{rel:R-f}
\end{equation}
where $\alpha$ is a constant of integration, and plugging the
relation~(\ref{rel:R-f}) into Eq.~(\ref{eq:met}), one finds
\begin{equation}
\nabla_\mu \nabla_\nu f - \frac12 g_{\mu\nu} \left[ f^3 - \alpha f \right] = 0. \label{eq:f}
\end{equation}

%%  Trivial solution
If $\partial_\mu f=0$, then Eq.~(\ref{eq:f}) has three trivial
solutions, $f=0$ and $f=\pm\sqrt\alpha$, and the corresponding
curvatures are given by $R=\alpha$ and $R=-2\alpha$, respectively.  In
the first solution, $\alpha$ can be arbitrarily chosen; it can be
positive, negative, or zero so that the model describes de Sitter
(dS), anti-de Sitter (AdS), or flat spacetime, respectively.  However,
in the second degenerate solutions, it should be definitely positive
so that the model describes only AdS spacetime.

%%  Cosmological solutions
To obtain a cosmological solution, we consider the metric ansatz,
$ds^2 = -d\tau^2 + a^2(\tau) dr^2$, where $a(\tau)$ is a scale factor.
Then, the first trivial solution, $f=0$, reads $R = 2\ddot{a}/a =
\alpha$, where the overdot~($\dot{\phantom{a}}$) denotes the
derivative with respect to $\tau$, so that the solution has the form,
\begin{equation}
\label{sol:triv1}
a(\tau) = 
\left\{
\begin{array}{ll}
  a_0 \exp \left[ \sqrt{\alpha/2} (\tau-\tau_0) \right], & \textrm{for } \alpha>0, \\
  a_0 + a_1 (\tau-\tau_0), & \textrm{for } \alpha=0, \\
  a_0 \left| \cos \left[ \sqrt{|\alpha|/2} (\tau-\tau_0) \right] \right|, & \textrm{for } \alpha<0.
\end{array}
\right.
\end{equation}
Note that the solution describes the exponentially expanding universe
known as dS ($R=\alpha>0$) cosmology as well as the oscillating
universe as an AdS ($R=\alpha<0$) one.  In addition, for $\alpha=0$,
it can also describe the steady state universe ($a_1=0$) as well as
uniformly expanding universe with constant expansion rate $a_1>0$.
%
%%  Trivial solution, $f=\pm\sqrt\alpha$
Next, the second degenerate trivial solutions, $f=\pm\sqrt\alpha$
($\alpha>0$), lead to $R = -2\alpha$, which means that the form of
scale factor is given by the oscillating AdS universe, as seen in the
last case of the first trivial solution~(\ref{sol:triv1}), $a(\tau) =
a_0 \left| \cos \left[ \sqrt{\alpha} (\tau-\tau_0) \right] \right|$.

%%  Non-trivial solution
On general ground, the relation~(\ref{rel:R-f}) yields $f=f(\tau)$ so
that the non-trivial components of Eq.~(\ref{eq:f}) are given by
\begin{align}
& \ddot{f} + \frac{1}{2} f \left[ f^2 - \alpha \right] = 0, \label{eq:non1} \\
& - a \dot{a} \dot{f} - \frac{1}{2} a^2 f \left[ f^2 - \alpha \right] = 0. \label{eq:non2}
\end{align}
Multiplied by $2\dot{f}$, Eq.~(\ref{eq:non1}) becomes
\begin{equation}
\label{eq:fdot}
\dot{f}^2 + \frac{1}{4} \left[ f^2 - \alpha \right]^2 = \gamma^2,
\end{equation}
where $\gamma$ is an integration constant. Note that when
$\alpha\le0$, $\gamma$ is restricted by $\gamma^2\ge\alpha^2/4$. If we
choose $\gamma^2=\alpha^2/4$, then $f=\dot{f}=0$ and the trivial
solution (of the first kind)~(\ref{sol:triv1}) is given.  Similarly,
for $\alpha>0$, if we choose $\gamma=0$, which means that $\dot{f}=0$
and $f=\pm\sqrt\alpha$, then we get the trivial solution of the second
kind.

\begin{figure}[pbt]
\begin{center}
  \includegraphics[width=0.35\textwidth]{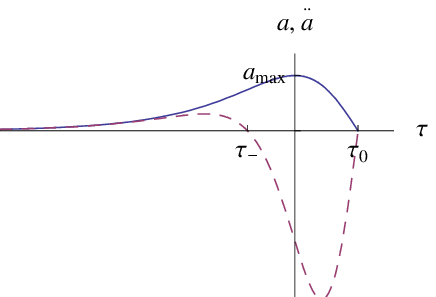}
  \includegraphics[width=0.35\textwidth]{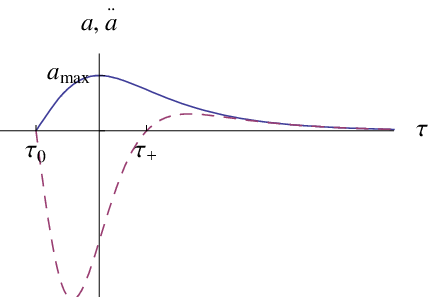}
  \caption{\label{fig:non}
    Two nontrivial solutions are shown.  They describe the phase
    changing from the accelerated universe to decelerated one and vice
    versa.  The solid curve represents the scale factor $a$ and the
    dashed curve denotes the acceleration $\ddot{a}$.  The transition
    time $\tau_\pm$ is given by $\tau_\pm = \tau_0 \pm
    \sqrt{2/\alpha}\ \mathrm{sech}^{-1} (1/\sqrt6)$.}
\end{center}
\end{figure}

Apart from these trivial solutions, for $\gamma^2 =
\frac{\alpha^2}{4}$ and $\alpha>0$, Eq.~(\ref{eq:fdot}) can be solved
as
\begin{equation}
\label{sol:f}
f = \xi \sqrt{2\alpha}\ \mathrm{sech} \left[ \sqrt{\frac{\alpha}{2}} (\tau-\tau_0) \right],
\end{equation}
where $\xi=\mathrm{sgn}(f)$ is the sign of $f$, i.e. $1$ for $f>0$ and
$-1$ for $f<0$.
Now, plugging Eq.~(\ref{eq:non2}) into Eq.~(\ref{eq:non1}) leads to
$\zeta \dot{f} = \beta a(\tau)$, where $\beta>0$ is an integration
constant and $\zeta=\mathrm{sgn}(\dot{f})$ so that the scale $a(\tau)$
is positive.  Then, from the solution~(\ref{sol:f}), the scale factor
is given by,
\begin{equation}
\label{scale}
a(\tau) = -\zeta \xi \frac{\alpha}{\beta}\ \mathrm{sech} \left[ \sqrt{\frac{\alpha}{2}} (\tau-\tau_0) \right] \mathrm{tanh} \left[ \sqrt{\frac{\alpha}{2}} (\tau-\tau_0) \right],
\end{equation}
which satisfies the relation~(\ref{rel:R-f}) as well.  The scale
factor~(\ref{scale}) describes a well-defined universe either in
$\tau\in(-\infty,\tau_0)$ for $\zeta\xi>0$ or in $\tau\in(\tau_0,\infty)$
for $\zeta\xi<0$.  Both universes end up with big crunches (see
Fig.~\ref{fig:non}).
%%%%%%%%%%%%%%%%%%%%%%%%%%%%%%%%%%%
%%  Figure: nontrivial solution  %%
%%%%%%%%%%%%%%%%%%%%%%%%%%%%%%%%%%%
However, no singularity appears in both universes, since the
relation~(\ref{rel:R-f}) tells us that
\begin{equation}
R = \alpha - 3 f^2 = \alpha - 6\alpha\ \mathrm{sech}^2 \left[ \sqrt{\frac{\alpha}{2}} (\tau-\tau_0) \right],
\end{equation}
which is finite for all $\tau$.  It is interesting to note that the
universe described by the solution~(\ref{scale}) has the transition
from dS to AdS or vice versa, since $R\to\alpha$ when
$\tau\to\pm\infty$ and $R\to-5\alpha$ when $\tau\to\tau_0$ without
recourse to some modification of equations of motion. 
It means that 
the smooth phase changing from the decelerated expansion of universe
to the accelerated one appears naturally.
This kind of phase transition has been discussed in the two-dimensional dilaton
gravity by assuming noncommutative Poisson algebra which modifies the
conventional equation of motion \cite{ky,ks}.  

Due to the
complicatedness, we have not presented other solutions, so that it
would be nice to find a better solution more closely to meet the
cosmological scenario.
In more realistic four-dimensional gravity with Chern-Simons modifications, it has been shown that two polarizations of a gravity wave carry intensities that are suppressed or enhanced by the modification~\cite{jp} and that elliptically polarized gravitational waves that are produced during inflation with $CP$-odd component can eventually generate the cosmic matter-antimatter asymmetry~\cite{aps}.

%%%%%%%%%%%%%%%%%%%%%%%
%%  Acknowledgments  %%
%%%%%%%%%%%%%%%%%%%%%%%

\begin{acknowledgments}
This work was supported by the Korea Science and Engineering Foundation 
(KOSEF) grant funded by the Korea government(MOST) 
(R01-2007-000-20062-0).
\end{acknowledgments}

%%%%%%%%%%%%%%%%%%
%%  References  %%
%%%%%%%%%%%%%%%%%%

\end{document}